\documentstyle[preprint,aps]{revtex}

\begin{document}

\begin{flushright}
  {\bf BU/HEPP/96-02} \\
 June 1996 \\
\end{flushright}
\vspace{1.cm}
\begin{center}

{\bf {\LARGE Lattice Charge Overlap II: Aspects of Charged Pion
Polarizability}}

\vspace{1.cm}

{\bf Walter Wilcox} \\

{\it Department of Physics, Baylor University, Waco, TX 76798-7316}

\end{center}

\begin{abstract}

Formulas are developed for use in lattice studies
of charged hadron polarizabilities. In particular, the valence quark
different-flavor component of the charged pion polarizability
is examined on a $16^{3}\times 24$ lattice at $\beta=6.0$ using Wilson fermions.
Using the elastic limit results of Part I of this series, it is concluded that this
represents a small negative component of the full charged polarizabilty.

\vskip\baselineskip

\end{abstract}

\vfill

\section{Introduction}

In the previous paper of this series, referred to here as Part I, the 
elastic limit of pions and rho mesons and were examined and the
results were compared to lattice simulations using three-point
function techniques. The elastic \lq\lq baseline" set in Part I makes
the measurement of nonelastic properties of these mesons possible.

Neutral hadron polarizabilities have been examined previously using the
techniques of lattice QCD\cite{fiebig}. The authors there used an external
field method to study the effect of a uniform electric field on the
correlation functions of mesons and baryons. However, as of yet
there have been no investigations of charged hadron polarizabilities
using the lattice techniques. This is unfortunate since
there is now a growing body of experimental information on
this subject. Of course, external field methods would be very difficult or
impossible to use in the case of charged hadrons because charged particles 
accelerate in an electric field. The definition of charged hadron
polarizability takes the form of the response of the particle to
an applied oscillating electromagnetic field. For this purpose a single
photon will do and the effect being considered is then just Compton
scattering.

The subject of charged pion polarizability will be considered from
the point of view of an effective relativistic field theory. Comparing
the general and phenomenological forms of the Compton scattering
amplitude to second order in the photon momentum will give us the form of
the polarizability coefficient. Next, the method for
extracting polarizability on the lattice is explained. It is pointed
out that the full pion polarizabilty requires the separate
evaluation of five types of connected and disconnected fermion
diagrams. In the fourth section the details of of our numerical simulation of the
valence different-flavor component of the charged pion polarizabilty will be given
and the results from Part I will be extensively used. It
will be established that the resulting contribution is a small negative component
of the full charged polarizabilty.
 
\section{Theory}

Charged hadron polarizability has been examined previously in Ref.\cite{grandy}
using the methods of charge overlap in the context of lattice QCD simulations.
However, the methods of Ref.\cite{grandy} are coordinate-space based;
the present techniques\cite{someofmine} are momentum-space based. First, let us
review the definition of charged pion polarizability as measured in Compton
scattering. The definition of charged polarizability
in terms of relativistic matrix elements in the context of an
effective meson theory will then be established.

\subsection{Review}

The general form of the charged pion Compton scattering amplitude,
\begin{equation}
T_{\mu \nu}= i\int d^{4}x e^{ik_{2}\cdot
x}(\pi(p_{2})|T(j_{\mu}(x)j_{\nu}(0))|\pi(p_{1})),
\label{Compton}
\end{equation}
where \lq $T$' denotes time ordering and
\begin{equation}
j_{\mu}(x)= q_{u}j_{\mu}^{u}(x)+q_{d}j_{\mu}^{d}(x),
\end{equation}
($q_{u}=\frac{2}{3}, q_{d}=-\frac{1}{3}$) is an 
electromagnetic current density for
up and down quarks, to second order in the photon momentum is given
by\cite{donoghue,terentev}
\begin{eqnarray} \nonumber
\sqrt{2E_{1}2E_{2}}T_{\mu \nu}=
-\frac{T_{\mu}(p_{1}+k_{1},p_{1})T_{\nu}(p_{2},p_{2}+k_{2})}{(p_{1}+k_{1})^{2}
-m^{2}}-\frac{T_{\mu}(p_{2},p_{2}-k_{1})T_{\nu}(p_{1}-k_{2},p_{1})}
{(p_{1}-k_{2})^{2}-m^{2}} \\
+2g_{\mu \nu} + A(k_{1}^{2}g_{\mu \nu}-k_{1\mu}k_{1\mu}
+k_{2}^{2}g_{\mu \nu}-k_{2\mu}k_{2\mu}) + B(k_{1}\cdot k_{2}g_{\mu
\nu}-k_{2\mu}k_{1\nu}) + Ct_{\mu\nu}.
\label{phenom}
\end{eqnarray}
In this context $A$, $B$ and $C$ are simply numerical coefficients.
A noncovariant continuum state normalization,
\begin{equation}
\sum_{n}\int \frac{d^{3}p}{(2\pi)^{3}} |n(p))(n(p)|={\bf 1},\label{normal}
\end{equation}
is being used which is responsible for the square root factor involving pion
energies appearing on the left hand side of Eq.(\ref{phenom}). The pion 
electromagnetic vertex ($q_{\mu}\equiv p_{\mu}'-p_{\mu}$),
\begin{equation}
T_{\mu}(p',p)=
(p_{\mu}'+p_{\mu})F_{\pi}(q^{2})+q_{\mu}\frac{p'^{2}-p^{2}}
{q^{2}}(1-F_{\pi}(q^{2})),
\label{vertex}
\end{equation}
has been written so as to obey
\begin{equation}
q_{\mu}T_{\mu}(p',p)=p'^{2}-p^{2},
\end{equation}
for off-shell pions, which is needed in order to satisfy the 
scalar electromagnetic Ward identity\cite{terentev}. 
To the fourth order in momentum (needed later),
\begin{equation}
F_{\pi}(q^{2})=1-\frac{<r^{2}>}{6}q^{2}+\frac{<r^{4}>}{120}q^{4}.
\label{expand}
\end{equation}
In the above $m$ is the pion mass and the tensor $t_{\mu\nu}$ is given by
\begin{equation}
t_{\mu\nu}= (k_{1}\cdot k_{2})Q_{\mu}Q_{\nu}+(Q\cdot k_{1})(Q\cdot k_{2})g_{\mu
\nu}-(Q\cdot k_{2})Q_{\mu}k_{1\nu}-(Q\cdot k_{1})Q_{\nu}k_{2\mu}.
\label{littlet}
\end{equation}
($Q_{\mu}=p_{1\mu}+p_{2\mu}$.) The Compton amplitude is conserved
($k_{1}^{\mu}T_{\mu\nu}=k_{2}^{\nu}T_{\mu
\nu}=0$), which can be shown to result in $A=\frac{<r^{2}>}{3}$ to this
order. The first two terms on the right of Eq.(\ref{phenom}) represent elastic
scattering from the pion while the others are contact terms.
(A metric with $g^{00}=1$, $g^{ii}=-1$, ($i=1,2,3$) is being used.) In order to
define electric and magnetic polarizabilities, let us examine the Compton
scattering amplitude $\alpha \epsilon^{\mu}_{1} T_{\mu\nu}
\epsilon^{\nu *}_{2}$, where $\alpha$ is the fine structure constant and
$\epsilon^{\nu}_{1}$ and $\epsilon^{\nu *}_{2}$ are the initial and final photon
polarization vectors, respectively. In the lab 
frame where the inital pion is at rest:
\begin{eqnarray}
k_{1}\cdot \epsilon_{1} =k_{2}\cdot \epsilon_{2}^{*} = 0,\\
p_{1}\cdot \epsilon_{1} =p_{1}\cdot \epsilon_{2}^{*} = 0.
\end{eqnarray}
Choose $\epsilon_{1}$ and $\epsilon_{2}^{*}$ to have only spatial components.
Defining $k_{(1,2)}^{0}=\omega_{(1,2)}$, in the nonrelativistic limit
($|\vec{k}_{(1,2)}|,\omega_{(1,2)}<<m$), one can show for off-shell photons that
\begin{equation}
\alpha \epsilon^{\mu}_{1} T_{\mu\nu}
\epsilon^{\nu *}_{2} = \hat{\epsilon}_{1} \cdot \hat{\epsilon}_{2}^{*}
[-\frac{\alpha}{m}(1+\frac{<r^{2}>}{6}(k_{1}^{2}+k_{2}^{2}))+\alpha_{E}
\omega_{1}\omega_{2}]+\beta_{M}(\hat{\epsilon}_{1}\times
\vec{k}_{1})\cdot (\hat{\epsilon}_{2}^{*}\times \vec{k}_{2}),\label{Camp}
\end{equation}
where
\begin{eqnarray}
\alpha_{E}\equiv -\alpha (\frac{B}{2m}+2mC),\\
\beta_{M}\equiv \alpha \frac{B}{2m},
\end{eqnarray}
are by definition the electric and magnetic polarizabilities, respectively.
An explicit form for $\alpha_{E}$ will now be exhibited and an
explanation of how this coefficient can be measured in a lattice simulation
will be given.

\subsection{Derivation}

First of all, let us use the continuum to lattice corespondences,
\begin{eqnarray}
|n(p)) \longrightarrow V^{1/2}|n(p)>,\\
j_{\mu}(x) \longrightarrow \frac{j_{\mu}^{L}(x)}{a^{3}},\label{currents}\\
\int d^{4}x \longrightarrow Va\int_{-\infty}^{\infty} dt\sum_{\vec{x}}
\end{eqnarray}
where $V=N_{s}a^{3}$, $N_{s}$ being the number of spatial 
sites in the lattice and
\lq a' being the spacing, to rewrite Eq.(\ref{Compton}) as
\begin{equation}
T_{\mu \nu}= iN_{s}^{2}a\int dt \sum_{\vec{x}} e^{ik_{2}\cdot
x}<p_{2}|T(j_{\mu}^{L}(x)j_{\nu}^{L}(0))|p_{1}>,
\label{compton2}
\end{equation}
Notice that time remains continuous (and is dimensionless) in this formalism and
that a possible finite renormalization factor has been neglected in
Eq.(\ref{currents}) since the exactly conserved lattice
charge density will be used here.

Next, a point about the amplitude being calculated needs to be made. On the
lattice, the appropriate object to consider is actually a normal ordered product of
currents:
\begin{equation}
T_{\mu \nu}^{eff}= iN_{s}^{2}a\int dt \sum_{\vec{x}} e^{ik_{2}\cdot
x}<\pi(p_{2})|:j_{\mu}^{L}(x)j_{\nu}^{L}(0):|\pi(p_{1})>.
\label{compton3}
\end{equation}
%
%
%
This is because the vacuum amplitude lattice Green function used
(see Eq.(\ref{amp}) below) contains a disconnected part when $p_{2}=p_{1}$
which must be removed by hand. The form of the disconnected
amplitude being subtracted in Eq.(\ref{compton3}) will later be exhibited
explicitly.

The methods of charge overlap will be used to calculate the necessary
amplitudes. Thus, we consider the four-point function corresponding to
$T_{00}$ with the kinematical conditions:
\begin{eqnarray}
\omega_{1}=\omega_{2}=0;~~p_{10}=p_{20}=m;~~|\vec{k}_{1}|=|\vec{k}_{2}|<<m.
\end{eqnarray}
The relation between
normal-ordered and time-ordered currents is
\begin{equation}
:j_{\mu}^{L}(x)j_{\nu}^{L}(0):~ = T(j_{\mu}^{L}(x)j_{\nu}^{L}(0))
~-<0|T(j_{\mu}^{L}(x)j_{\nu}^{L}(0))|0>.
\end{equation}
Using lattice completeness,
\begin{equation}
\sum_{n,\vec{p}_{n}}|n(\vec{p}_{n})><n(\vec{p}_{n})|={\bf 1},
\end{equation}
as well as space-time translations,
\begin{equation}
j_{\mu}^{L}(x)= e^{ip\cdot x}j_{\mu}^{L}(0)e^{-ip\cdot x},
\end{equation}
the space sum and time integral in Eq.(\ref{compton3}) now results in
\begin{eqnarray}
T_{00}^{eff}|_{\lq\lq kinematics"}=
2N_{s}^{2}\sum_{n}\frac{|<\pi(\vec{0})|\rho^{L}(0)|n(\vec{k})>|^{2}}{(E_{n}-m)}
-2\frac{N_{s}}{a^{3}}\sum_{n}\frac{|<0|\rho^{L}(0) |n(\vec{k})>|^{2}}{E_{n}}.
\label{effective}
\end{eqnarray}

Contact between Eq.(\ref{effective}) and Eq.(\ref{phenom}) above
now needs to be made. Separate off the elastic part in Eq.(\ref{effective}),
\begin{eqnarray}
T_{00}^{el}\equiv
2N_{s}^{2}\frac{|<\pi (\vec{0})|\rho^{L}(0)|\pi (\vec{k})>|^{2}}{(E_{\pi}-m)},
\end{eqnarray}
\begin{eqnarray}
<\pi (\vec{0})|\rho^{L}(0)|\pi (\vec{k})>=
\frac{1}{N_{s}}\frac{(E_{\pi}+m)}{\sqrt{2E_{\pi}2m}}F_{\pi}(q^{2}),
\end{eqnarray}
where $F_{\pi}(q^{2})$ is the usual continuum pion form factor. Using
Eq.(\ref{expand}), one can show after some algebra that
\begin{equation}
T_{00}^{el}=
\frac{4m}{\vec{k}^{2}}+\{ \frac{1}{m} -\frac{4}{3}m<r^{2}> \} +\vec{k}^{2} \{
-\frac{<r^{2}>}{3m} +\frac{1}{9}m<r^{2}>^{2}+\frac{1}{15}m<r^{4}>
\},\label{elastic}
\end{equation}
up to second order in the photon momentum. For the nonelastic part one may show
at low momentum ($|\vec{k}|<<m$) for vector or axial vector states
that
\begin{eqnarray}
<\pi (\vec{0})|\rho^{L}(0)|n (\vec{k})>=\frac{\epsilon^{0}(k)d_{\pi
n}m_{n}}{N_{s}},\label{expand1}\\ <0|\rho^{L}(0)|n
(\vec{k})>=\frac{\epsilon^{0}(k)d_{0 n}m_{n}a^{1/2}}{\sqrt{N_{s}}},
\label{expand2}
\end{eqnarray}
which serves to define the constants $d_{\pi n}$ and $d_{0 n}$, and
$\epsilon^{0}(k)$ is the 0th component of a polarization vector for the state
$|n(\vec{k})>$ with mass $m_{n}$. The $d_{\pi n}$, $d_{\pi 0}$ factors are defined
in these expressions so that they are independent of the lattice size and
spacing. Interestingly, vectors and axial vectors are the only types of
intermediate states which may contribute to the polarizability. See the Appendix
for further details. If we choose a helicity basis where the
${\epsilon}^{\mu}(k,\lambda)$ with $\lambda =+,-$ are purely spatial and
the positive-z axis is along $\vec{k}$, then only the longitudinal
polarization state, $\lambda=0$, contributes in Eqs.(\ref{expand1}) and
(\ref{expand2}),
\begin{equation}
\epsilon^{0}(k,0)=\frac{|\vec{k}|}{m_{n}},\label{long}
\end{equation}
and the expansion to second order in photon momentum for these
kinematics then yields
\begin{eqnarray}
T_{00}^{eff}|_{\lq\lq kinematics"}=T_{00}^{el}+2\vec{k}^{2}(\sum_{n\ne
\pi}\frac{|d_{\pi n}|^{2}}{(E_{n}-m)} -V \sum_{n}\frac{|d_{0
n}|^{2}}{E_{n}}),\label{finaleff}
\end{eqnarray}
where $T_{00}^{el}$ is given by Eq.(\ref{elastic}).

The phenomenological expression, Eq.(\ref{finaleff}), is to be compared with
the $\mu=\nu=0$ component of the general Compton amplitude,
Eq.(\ref{phenom}), also to order $\vec{k}^{2}$ with the same kinematics. One may
show that
\begin{eqnarray}
2mT_{00}|_{\lq\lq kinematics"}&=&
\frac{T_{0}(p_{1}+k_{1},p_{1})T_{0}(p_{2},p_{2}+k_{2})}{(p_{1}+k_{1})^{2}
-m^{2}}-\frac{T_{0}(p_{2},p_{2}-k_{1})T_{0}(p_{1}-k_{2},p_{1})}
{(p_{1}-k_{2})^{2}-m^{2}}\nonumber \\ & & \mbox{} +2
-\frac{2<r^{2}>}{3}\vec{k}^{2}-B\vec{k}^{2} -4m^{2}\vec{k}^{2}C.
\end{eqnarray}
For these kinematics, one has
\begin{eqnarray}
\frac{1}{(p_{1}+k_{1})^{2}
-m^{2}}=-\frac{1}{(p_{1}-k_{2})^{2}-m^{2}}=-\frac{1}{\vec{k}^{2}},
\end{eqnarray}
and also, for example,
\begin{eqnarray}
T_{0}(p_{1}+k_{1},p_{1})=2m(1-\frac{<r^{2}>}{6}q^{2}+\frac{<r^{4}>}{120}q^{4}).
\end{eqnarray}
When the pieces are put together, one obtains
\begin{eqnarray}
T_{00}|_{\lq\lq kinematics"}&=&\frac{4m}{\vec{k}^{2}}+\{ \frac{1}{m}
-\frac{4}{3}m<r^{2}> \}  +\vec{k}^{2} \{ -\frac{<r^{2}>}{3m}
+\frac{1}{9}m<r^{2}>^{2}\nonumber \\ & &\mbox{}+\frac{1}{15}m<r^{4}> -
(\frac{B}{2m}+2mC) \}.\label{finalphen}
\end{eqnarray}
As the last step in the derivation, compare Eq.(\ref{finaleff}) and
Eq.(\ref{finalphen}) and identify
\begin{eqnarray}
\alpha_{E}=-\alpha(\frac{B}{2m}+2mC)=2\alpha\{ \sum_{n\ne \pi}\frac{|d_{\pi
n}|^{2}}{(E_{n}-m)} -V \sum_{n}\frac{|d_{0 n}|^{2}}{E_{n}} \}.
\label{formula}
\end{eqnarray}

This formula relates $\alpha_{E}$ to properties of relativistic matrix elements,
which was the goal. Of course, the $d_{\pi n}$ and $d_{0 n}$ factors are nothing
but appropriate derivatives of the matrix elements in Eq.(\ref{expand2}) with
respect to $|\vec{k}|$, evaluated at $\vec{k}=0$. There are several other aspects
of this formula worthy of mention. First, the explicit appearance of a
volume factor makes it clear that neither term above is finite, but the
whole expression must be regulated (as on the lattice) to extract a finite
answer. Second, it does not have the classical charge radius term,
$\alpha$$<r^{2}>$$/3m$. This is because it comes from comparing to an
effective theory of point mesons. (Even the pion is viewed as a point object in
this case; it's \lq\lq size" simply comes from additional local interactions with
photons.) Third, it explicitly includes the vacuum polarization contribution first
explained by Terent'ev in Ref.\cite{terentev2}. The negative sign in front of this
term makes it abundantly clear why in relativistic field theories the charged
particle's polarizability can be positive or negative. The polarizability of the
pion, no longer considered the groundstate as in nonrelativistic quantum
mechanics, is actually defined relative to the vacuum, a result remininiscent of
the Casimir effect. Finally, even though we are only keeping terms to order
$\vec{k}^{2}$, it is clear that the above is an exact formula for the
polarizability, which is simply the numerical coefficient of the second order
energy term in Eq.(\ref{Camp}).

\section{Lattice Measurement}

Let us concentrate now on the lattice amplitudes necessary to directly extract
the charged pion polarizability. Some comments about
indirect lattice measurements of this quantity will be made later.

For these purposes, one forms the usual charge overlap function, but
with normal ordering of the charge densities instead of time ordering:
\begin{equation}
P_{\pi}(\vec{r};t_{3},t_{2},t_{1})\equiv
\frac{\sum_{\vec{x}_{3},\vec{z}}<0|\phi^{\dagger}(x_{3}):\rho^{L}(x_{2})
\rho^{L}(x_{1}):\phi(z)|0>}{\sum_{\vec{x}_{3},\vec{z}}<0|
\phi^{\dagger}(x_{3})\phi(z)|0>},\label{amp}
\end{equation}
where $t_{3}>(t_{2},t_{1})>0$, $z=(0,\vec{z})$
and  $r\equiv x_{2}-x_{1}$. Dimensionless times and
distances are being used in this section. The formulas are now in discrete
Euclidean space but one imagines the time axis is still continuous. 
The time limits $t_{3}>>t_{1,2}>>1$ yield
\begin{eqnarray}
P_{\pi}(\vec{r};t_{3},t_{2},t_{1})\longrightarrow 
<\pi(\vec{0})|:\rho^{L}(x_{2})
\rho^{L}(x_{1}):|\pi(\vec{0})>, 
\end{eqnarray}
where
\begin{eqnarray} 
<\pi(\vec{0})|:\rho^{L}(x_{2})
\rho^{L}(x_{1}):|\pi(\vec{0})>&=& <\pi(\vec{0})|T(\rho^{L}(r)
\rho^{L}0))|\pi(\vec{0})>\nonumber \\ & &\mbox{} -<0|T(\rho^{L}(r)
\rho^{L}(0))|0>.
\end{eqnarray}

Consider now the finite Fourier transform of
$P_{\pi}(\vec{r};t_{3},t_{2},t_{1})$, 
\begin{eqnarray}
Q(\vec{q};t_{3},t_{2},t_{1})\equiv N_{s}\sum_{\vec{r}}e^{-i\vec{q}\cdot
\vec{r}}P_{\pi}(\vec{r};t_{3},t_{2},t_{1}).
\end{eqnarray}
In the same time limit:
\begin{eqnarray}
Q(\vec{q};t_{3},t_{2},t_{1})
\longrightarrow N_{s}\sum_{\vec{r}}e^{-i\vec{q}\cdot
\vec{r}}\{<\pi(\vec{0})|T(\rho^{L}(r)
\rho^{L}0))|\pi(\vec{0})>
-<0|T(\rho^{L}(r)
\rho^{L}(0))|0>\}.
\end{eqnarray}
Assuming $t_{2}-t_{1}>0$ and inserting a complete set of states now gives:
\begin{eqnarray}
Q(\vec{q};t_{3},t_{2},t_{1})
&\longrightarrow& N_{s}^{2}\{ \sum_{n}|<\pi(\vec{0})|\rho^{L}(0)
|n(\vec{q})>|^{2}e^{-(E_{n}a-m_{\pi}a)(t_{2}-t_{1})} \nonumber \\
& &\mbox{} -\sum_{n}|<0|\rho^{L}(0)|n(\vec{q})>|^{2}e^{-E_{n}a(t_{2}-t_{1})} \}.
\end{eqnarray}
Separate off the elastic term ($n=\pi$),
\begin{eqnarray}
Q^{el}(\vec{q};t_{2}-t_{1})
\equiv N_{s}^{2}|<\pi(\vec{0})|\rho^{L}(0)
|\pi (\vec{q})>|^{2}e^{-(E_{\pi}a-m_{\pi}a)(t_{2}-t_{1})},
\end{eqnarray}
above. Then, using Eqs.(\ref{expand1}), (\ref{expand2}) and (\ref{long}) at low
momentum transfer results in the construction,
\begin{eqnarray}
\alpha a\int_{-\infty}^{\infty} d(t_{2}-t_{1})
[Q(\vec{q};t_{3},t_{2},t_{1})-Q^{el}(\vec{q};t_{2}-t_{1})]\nonumber \\
= 2\alpha \vec{k}^{2}(\sum_{n\ne \pi}\frac{|d_{\pi
n}|^{2}}{(E_{n}-m)} -V \sum_{n}\frac{|d_{0 n}|^{2}}{E_{n}})
= \vec{k}^{2}\alpha_{E},\label{lattice}
\end{eqnarray}
($\vec{k}=\vec{q}/a$) when the relative time integral between the two charge
densities is explicitly done. (One gets a doubling of the result when the other
time relationship, $t_{2}-t_{1}<0$, is assumed.) Of course on a finite lattice the
time integrals do not really extend to $\pm \infty$ because of the fixed time
locations of the particle sources, but instead to time limits such
that the elastic limit is effectively established. Eq.(\ref{lattice}) now gives us
the means of measuring the Compton scattering pion polarizability coefficient. 
On the lattice one may take a numerical \lq\lq derivative" of the left hand side
of Eq.(\ref{lattice}) with respect to $\vec{q}^{\,2}$ by simply dividing it
by $\vec{q}^{\,2}$at the lowest lattice momentum. (The relative error in this
procedure is of order $\vec{q}^{\,2}$.) 
 
It is crucial to work in momentum space to extract the
polarizability. In position space one would look at (image corrections are
assumed), 
\begin{eqnarray}
R^{2}(t_{3},t_{2},t_{1}) \equiv 
\sum_{\vec{r}}\vec{r}^{\,2}P_{\pi}(\vec{r};t_{3},t_{2},t_{1}),
\end{eqnarray}
which does not project onto good momentum when a complete set of
states is inserted\cite{grandy}. Thus, the limitation to small
momenta, necessary to use Eqs.(\ref{expand1}) and (\ref{expand2}) above,
can not be enforced.

Although Eq.(\ref{lattice}) is a simple formula, it's direct
evaluation on the lattice via four-point functions which arise from the field
contractions in Eq.(\ref{amp}) is not easy. There are five
types of diagrams, which may be classified by fermion line topology, as shown in
Fig.1. In the case of Figs.1(a) and (b), the photons interact
directly with the valence quarks. Fig.1(a) will be termed the same-flavor valence
part and Fig.1(b) will be called the different-flavor valence part. Figs.1(c), (d)
and (e) represent contributions from disconnected fermion lines in which arbitrary
numbers of gluon lines connect the various pieces together. These gluon
connections are enforced in the correlations between the pion propagator and the
various fermion loops in the lattice Monte Carlo. (When isolated groups of fermion
lines arise, \lq\lq connected part" means \lq\lq statistically correlated part" on
the lattice.) Fig.1(c) represents the contribution from correlations between a
valence quark and a self-contracted quark loop. (Flavors other than $u$ and $d$
can of course contribute to the quark loops.) Fig.1(d) has both currents, either
same-flavor or different-flavor, self-contracted. In the case of Fig.1(e), one has
a complete quark loop formed. It is the amplitude where the pion propagator is
completely disconnected from the quark loops in (d) and (e) (when
$p_{2}=p_{1}$) which needs explicit subtraction and is responsible for the
negative contribution on the right hand side of
Eq.(\ref{formula}). Fig.1(c) also has a disconnected part, however, there is no
need to do a subtraction since it can be
shown that the vacuum expectation value of the exactly conserved lattice charge
density vanishes configuration by configuration for color SU(2), and in the
configuration average for SU(3).

The lattice formula for $\alpha_{E}$ has now been derived and the agenda for
evaluating the charged pion polarizability directly on the lattice has been set.
In the next section we will examine one particular type of diagram
and show that it actually makes only a small contribution to $\alpha_{E}$.

\section{Lattice Simulation}

As pointed out immediately above, the direct calculation of the charged pion
polarizability involves many diagrams which must be separately evaluated. In this
section the different-flavor valence piece,
Fig.1(a), will be considered. 

The full different-flavor piece involves Figs.1(a), (c) and (d) and can be
evaluated on the lattice at low lattice momentum, $\vec{q}$, by
\begin{eqnarray}
\alpha_{E}^{u,d}=\frac{\alpha a^{3}}{\vec{q}^{\,2}}\int_{-\infty}^{\infty}
d(t_{2}-t_{1}) [Q^{u,d}(\vec{q};t_{3},t_{2},t_{1})-Q^{el;u,d}
(\vec{q};t_{2}-t_{1})],
\end{eqnarray}
where $Q^{u,d}(\vec{q};t_{3},t_{2},t_{1})$ and $Q^{el;u,d}
(\vec{q};t_{2}-t_{1})$ are identified from the Fourier transform of
\begin{equation}
P_{\pi}^{u,d}(\vec{r};t_{3},t_{2},t_{1})\equiv 2q_{u}q_{d}
\frac{\sum_{\vec{x}_{3},\vec{z}}<0|\phi^{\dagger}(x_{3}):\rho^{L;u}(x_{2})
\rho^{L;d}(x_{1}):\phi(z)|0>}{\sum_{\vec{x}_{3},\vec{z}}<0|
\phi^{\dagger}(x_{3})\phi(z)|0>},\label{u,d}
\end{equation}
as above. 

We now specialize to evaluating the valence $u$, $d$ contribution from
Fig.1(a). We refer the reader to Part I for a discussion of the details of
the numerical simulation, which was carried out on $20$ quenched $16^{3}\times 24$
lattices with Wilson fermions at $\beta=6.0$. The
appropriate correlation functions for three quark masses are
displayed in Fig.5 of Part I. According to Eq.(\ref{lattice}), we now need to
evalute the area under these functions and subtract the elastic contribution. The
results of this procedure are presented in Table I, where the errors specified
are purely statistical. In arriving at these results, the numerical value $\alpha
a^{3}=1.06\times 10^{-44}$cm$^{3}$ has been used, where $\alpha$ is the fine
structure constant. This value is inferred from the inverse lattice spacing,
$a^{-1}=1740$ MeV, which is taken from comparing the chirally extrapolated lattice
nucleon mass with the experimental value\cite{scale}. 

When the results in Table I are extrapolated linearly in quark mass,
as defined in Part I, to the chiral limit at
$\kappa_{cr}=0.1564$, we obtain $(\alpha_{E}^{u,d})^{valence}=-.12 (11)\times
10^{-4}$ fm$^{3}$.

There are various systematic uncertainties in our final numerical result.
One systematic uncertainty is associated with the
numerical evaluation of the discrete nonelastic time correlation integral. 
This discrete time integral was done using both the trapezoidal rule and
Simpson's method\cite{nbs}. The differences
for all three $\kappa$ values were less than $2\%$. There is an additional
systematic uncertainty associated with the numerical momentum derivative. As
pointed out above, this is expected to be of order $\vec{q}^{\,2}$, or about
$10\%$, the same as other finite lattice spacing errors. This was examined
numerically by repeating the simulation (with the same fits) on
$10$ configurations, also at $\beta=6.0$, on a larger $20^{3}\times 30$
lattice at $\kappa=.152$. The result on the larger lattices is
$-.33 (14)\times 10^{-4}$ fm$^{3}$, which unfortunately has large statistical
errors, but is consistent with the value in Table I. In
addition to these uncertainties, there is also the systematic uncertainty
associated with the setting of the lattice scale. Since the pion polarizability
scales as $a^{3}$, it is rather sensitive to a change in the scale. In order to
remain consistent with the low energy lattice results for nucleon mass, pion decay
constant and proton electric form factor in Ref.\cite{scale}, a roughly $10\%$
change in the present lattice scale could be tolerated. This means a systematic
scale uncertainty of approximately $33\%$ for $\alpha_{E}$.

\section{Summary and Conclusions}

The establishment of the charge overlap elastic limit in Part I allows the
possibility of obtaining nonelastic hadron properties directly from
lattice QCD. In particular, using the lattice formulas developed,
we have see that the charged pion polarizability can be measured
on the lattice from low momentum charge overlap correlation functions.

We have obtained $(\alpha_{E}^{u,d})^{valence}=-.12 (11)\times
10^{-4}$ fm$^{3}$ in the chiral limit from the quark diagram
in Fig.1(a). This is to be compared with the experimental result\cite{mark2}
\begin{eqnarray}
\alpha_{E}^{exp} = 2.2 (16) \times 10^{-4} {\rm fm}^{3},\nonumber
\end{eqnarray}
and the result from second order chiral pertubation theory\cite{chiral},
\begin{eqnarray}
\alpha_{E}^{chiral} = 2.4 (5) \times 10^{-4} {\rm fm}^{3}.\nonumber
\end{eqnarray}
Thus, the present results establish that the valence quark different-flavor
diagram, Fig.1(a) contributes only a small, negative component to the full
charged pion polarizability. We must therefore look to the other 
diagrams in Fig.1 for the bulk of this effect.

It is important to note that in addition to the present, direct measurement
of the charged pion polarizability, there is also the Das, Mathur, Okubo (DMO)
current algebra sum rule\cite{das} for $\alpha_{E}$. Although this formula is not
exact, the measurements required for it's evaluation on the lattice are much
simpler than the direct method since it involves only mesonic two-point functions.
A lattice evaluation of the so-called intrinsic part of the charged pion
polarizability in the DMO sum rule should be quite interesting and is currently
underway.

Finally, it was noted above that the lattice measurement of $\alpha_{E}$
scales like $a^{3}$ and is therefore quite sensitive to the lattice scale. This
opens up the possibility of using charged pion polarizability as 
a place to accurately test QCD, or alternately, to set the lattice scale,
once the lattice and experimental results are both improved.

\section{Acknowledgments}

This work is supported in part by the NSF under Grant No.\ PHY-9401068. 
It is also supported by the National Center for Supercomputing
Applications (NCSA) at the University of Illinois at Urbana-Champaign
and used the CM5 and SGI Power Challenge computers. 

\newpage

\begin{center} {\bf Appendix}
\end{center}

In this Appendix it is established that the only intermediate states which may
contribute in Eq.(\ref{formula}) to the charged pion polarizability are vectors and
axial vectors.

First, let us dispose of the spin zero case. We have ($n\ne\pi$)
\begin{equation}
(\pi (p)|j_{\alpha}(0)|n (p_{n}))= \frac{p_{\alpha}G_{1}(q^{2})
+p_{n\alpha}G_{2}(q^{2})}{(2E_{\pi}2E_{n})^{1/2}}.\label{spin0}
\end{equation}
where as usual $q^{2}=(p-p_{n})^{2}$. Charge conservation requires
the right hand side of Eq.(\ref{spin0}) dotted into $p^{\mu}_{n}-p^{\mu}$ to
be zero, so that
\begin{equation}
G_{2}(q^{2})= G_{1}(q^{2})(\frac{m^{2}-p\cdot p_{n}}{m_{n}^{2}-p\cdot
p_{n}}).
\end{equation}
Therefore (considering the pion at zero spatial momentum)
\begin{equation}
(\pi (\vec{0})|j_{4}(0)|n (p_{n}))=\frac{G_{1}(q^{2})}{(2m2E_{n})^{1/2}}(m
+E_{n}\frac{m^{2}-mE_{n}}{m_{n}^{2}-mE_{n}}).
\end{equation}
At low momentum, $|\vec{p}_{n}|<<m$, we then find
\begin{equation}
(\pi (\vec{0})|j_{0}(0)|n (p_{n}))\propto G_{1}((m-m_{n})^{2})\vec{p}_{n}^{\,2}.
\end{equation}
Therefore the absolute square of this matrix element goes at least like
$\vec{p}_{n}^{\,4}$ at low momentum, assuming $G_{1}((m-m_{n})^{2})$ is
nonsingular, and can not contribute to the pion polarizability term in
$T_{00}^{eff}$, which is proportional to $\vec{p}_{n}^{\,2}$.

We now consider general nonzero spin, s. Some fundamental 
properties of polarization tensors are: 
($\lambda_{s}=-s,,,0,,,s$):
\begin{eqnarray}
\epsilon^{\mu_{1}\cdots \mu_{s}}(p,\lambda_{s})\epsilon_{\mu_{1}\cdots
\mu_{s}}^{*}(p,\lambda_{s}')&=&\delta_{\lambda_{s},\lambda_{s}'}, \\
g_{\mu_{\alpha}\mu_{\beta}}\epsilon^{\mu_{1}\cdots
\mu_{\alpha}\cdots \mu_{\beta}\cdots \mu_{s}}(p,\lambda_{s})&=&0,\label{fun1}\\
 p_{\mu_{\alpha}}\epsilon^{\mu_{1}\cdots
\mu_{\alpha}\cdots \mu_{s}}(p,\lambda_{s})&=&0\label{fun2}.
\end{eqnarray}

Adopting a helicity basis, the Lorentz structure for spin s must be of the
form
\begin{equation}
(\pi (p)|j_{\alpha}(0)|n
(s,\lambda_{s},p_{n}))=\frac{\epsilon^{\mu_{1}\cdots
\mu_{s}}(p_{n},\lambda_{s})A_{\mu_{1}\cdots
\mu_{s}\alpha}(p,p_{n})}{(2E_{\pi}2E_{n})^{1/2}}.\label{matrix}
\end{equation}
where it is assumed that the tensor $A_{\mu_{1}\cdots \mu_{s}\alpha}(p,p_{n})$ is
nonsingular. 

First, we will need the following result:
\begin{equation}
\left. \begin{array}{c}
\epsilon^{0\cdots 0}(p,\lambda_{s}) \propto |\vec{p}|^{s},~~\lambda_{s}= 0,
  \\  
\epsilon^{0\cdots 0}(p,\lambda_{s}) = 0, ~~\lambda_{s} \ne  0 .  
\end{array} \right\}\label{theorem}
\end{equation}
This can be established from the higher spin considerations in Ref.\cite{sch}.
This reference shows that the polarization vectors for higher spin may be
constructed from those of spin 1 via
\begin{equation}
\sum_{\lambda_{s}=-s}^{s}\psi_{s\lambda_{s}}(\xi)\epsilon^{\mu_{1}\cdots
\mu_{s}}(p,\lambda_{s})x_{\mu_{1}}\cdots x_{\mu_{s}}=\left[
\sum_{\lambda_{1}=-1}^{1}\psi_{1\lambda_{1}}(\xi)\epsilon^{\mu}(p,\lambda_{1})
x_{\mu} \right]^{s},\label{ugly}
\end{equation}
where
\begin{equation}
\psi_{s\lambda_{s}}(\xi)\equiv \left[
\frac{(2s)!}{(s+\lambda_{s})!(s-\lambda_{s})!}
\right]^{1/2}\xi_{+}^{s+\lambda_{s}}\xi_{-}^{s-\lambda_{s}}.
\end{equation}
Using a purely timelike $x_{\mu}$ and recalling that
$\epsilon^{0}(p,\lambda_{1})\ne 0$ only when $\lambda_{1}=0$, 
it is easy to see by matching coefficients of
$\xi_{+}$ and $\xi_{-}$ on both sides of Eq.\ref{ugly} that
\begin{equation}
\left. \begin{array}{c}
\epsilon^{0\cdots 0}(p,\lambda_{s}) \propto
(\epsilon^{0}(p,0))^{s},~~\lambda_{s}= 0,
  \\  
\epsilon^{0\cdots 0}(p,\lambda_{s}) = 0, ~~\lambda_{s} \ne  0 .  
\end{array} \right\} .
\end{equation}
Now using $\epsilon^{0}(p,0)\propto |\vec{p}|$ from Eq.(\ref{long}) establishes
Eq.(\ref{theorem}) above. There are many other ways of proving
this result. 
%
%
%
%
%

Working in the rest frame of the pion and considering only
$j_{0}$, the matrix element in Eq.(\ref{matrix}) is now seen
to be proportional to $\epsilon^{0\cdots 0}(p,\lambda_{s})$:
\begin{equation}
(\pi (\vec{0})|j_{0}(0)|n(s,\lambda_{s},p_{n}))\propto \epsilon^{0\cdots
0}(p_{n},\lambda_{s}).
\end{equation}
This comes about through repeated use of Eqs.(\ref{fun1}) and (\ref{fun2}).
For example, consider the $s=2$ case. Explicitly, one has
\begin{eqnarray}
(\pi(\vec{0})|j_{0}(0)|n(s,\lambda_{s},p_{n}))&=&\frac{1}{(2m2E_{n})^{1/2}}
\left( \epsilon^{i_{1}i_{2}}A_{i_{1}i_{2}0}
+\epsilon^{i_{1}0}A_{i_{1}00}\right.
\nonumber \\ & & \mbox{} \left.+ \epsilon^{0i_{2}}A_{0i_{2}0}+
\epsilon^{00}A_{000} \right).\label{explicit}
\end{eqnarray}
The remaining spatial $O(3)$ symmetry of the matrix element requires that
$A_{i_{1}00}\propto (p_{n})_{i_{1}}$, $A_{0i_{2}0}\propto (p_{n})_{i_{2}}$, and
$A_{i_{1}i_{2}0}\propto (p_{n})_{i_{1}}(p_{n})_{i_{2}}$ or $A_{i_{1}i_{2}0}\propto
g_{i_{1}i_{2}}$. Through the use of Eqs.(\ref{fun1}) and (\ref{fun2}) each
term in Eq.(\ref{explicit}) is seen to be proportional to
$\epsilon^{00}(p_{n},\lambda_{2})$. This in turn means that the matrix element
couples only to $\lambda_{2}=0$ and behaves like
$\vec{p}_{n}^{\,2}$ (or possibly higher power) at low momentum from
Eq.(\ref{theorem}). This argument can be extended to higher spin. Thus,
$s=1,\lambda_{1}=0$ (axial) is the only possible contributor to the charged pion
polarizability through this matrix element.

Although only the matrix element $(\pi(\vec{0})|j_{0}(0)|n(s,\lambda_{s},p_{n}))$
has been discussed, similar conclusions hold for
$(0|j_{0}(0)|n(s,\lambda_{s},p_{n}))$; namely, that the state
$n(s,\lambda_{s},p_{n})$ contributes to charged pion polarizability only when
$s=1, \lambda_{1}=0$ (vector).

\newpage

\begin{center}
{\bf Table} \\
\end{center}
\vspace{1.cm}

TABLE I. Valence contribution to the dfferent-flavor
charged pion polarizability as a function of $\kappa$.

\vspace{.3cm}
\begin{center}
\begin{tabular}{cc}\hline\hline
$\kappa$ & $\alpha_{E}$ (fm$^{3}$) \\
\hline
  $~~0.148~~$  & $-0.29 (6)\times 10^{-4}$ \\ 
  $~~0.152~~$  & $-0.23 (6)\times 10^{-4}$ \\ 
  $~~0.154~~$  & $-0.15 (10)\times 10^{-4}$ \\ \hline\hline
\end{tabular}
\end{center} 
\vfill
\newpage

\begin{center} {\bf Figure Caption}

\end{center}
     
\begin{enumerate}

\item Graphical representation of the four-point functions
contributing to charged pion polarization. The zero-momentum pion interpolating
fields are represented by horizontal bars, the time axis runs horizontally and the
vertical direction represents one space dimension. (a) and (b) represent
different-flavor and same-flavor valence contributions, respectively. Fig.1(c)
is the contribution from a single isolated quark loop, while (d) is the
two quark loop contribution. (e) arises from a same-flavor current-current
contraction. It is assumed that gluon lines connect the various fermion lines
together in each diagram. Diagrams (d) and (e) have a disconnected part which must
be subtracted.

\end{enumerate}
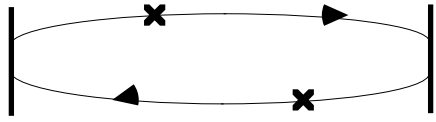
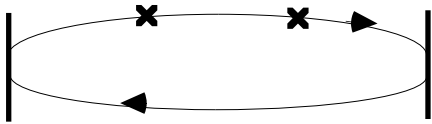
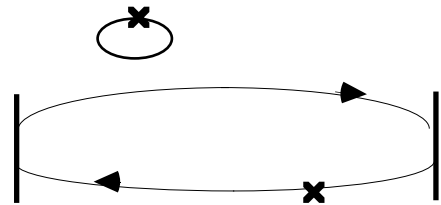
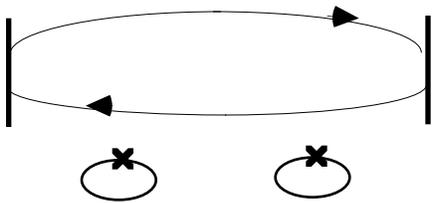
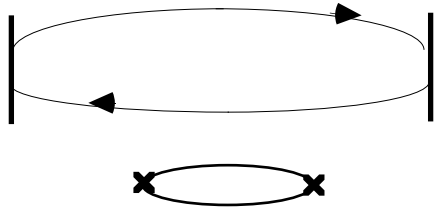
\end{document}